\documentclass[aps,prd,nofootinbib,groupedaddress,twocolumn]{revtex4}
\usepackage{graphicx}
\usepackage{epsfig}
\usepackage{dcolumn}
\usepackage{amsfonts}
\usepackage{amssymb}
\usepackage{bm}
\usepackage{amsmath}
\def\Journal#1#2#3#4{{#1} {\bf#2}, #3 (#4)}

\def\PLB{{\rm Phys. Lett.}  B}
\def\PRL{\rm Phys. Rev. Lett.}
\def\PRD{{\rm Phys. Rev.} D}

\def\ep{\epsilon}
\def\vep{\varepsilon}
\def\la{\langle}
\def\ra{\rangle}

\def\al{\alpha}

\def\be{\begin{equation}}
\def\ee{\end{equation}}
\def\bea{\begin{eqnarray}}
\def\eea{\end{eqnarray}}
\begin{document}
\title{Ideas of Four-Fermion Operators in Electromagnetic Form Factor Calculations}
\author{Chueng-Ryong Ji$^a$, Bernard L.G. Bakker$^b$, Ho-Meoyng Choi$^c$,
Alfredo Suzuki$^{a,d}$}
\thanks{Present address: Southern Adventist University, Collegedale, TN 37315, USA}
\affiliation{$^a$ Department of Physics, North Carolina State University,\\
Raleigh, NC 27695-8202,USA\\
$^b$Department of Physics and Astrophysics, Vrije Universiteit,\\
De Boelelaan 1081, NL-1081 HV Amsterdam, The Netherlands\\
$^c$ Department of Physics, Teachers College, Kyungpook National University,
Daegu, Korea 702-701\\
$^d$ Instituto de F\'{\i}sica Te\'orica-UNESP Universidade Estadual Paulista\\
Rua Dr. Bento Teobaldo Ferraz, 271 - Bloco II - 01140-070, S\~ao Paulo, SP,
Brazil }

\begin{abstract}
Four-fermion operators have been utilized in the past to link the
quark-exchange processes in the interaction of hadrons with the effective
meson-exchange amplitudes. In this paper, we apply the  similar idea
of a Fierz rearrangement to the electromagnetic processes and focus
on the electromagnetic form factors of the nucleon and the electron.
We explain the motivation of using four-fermion operators and discuss
the advantage of this method in computing electromagnetic processes.
\end{abstract}


\maketitle

\section{Introduction}
\label{sec.I}
\begin{figure}[htb]
\begin{center}
 \includegraphics[width=60mm]{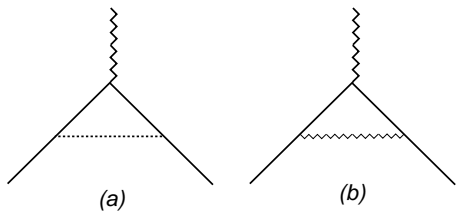}
\caption{Triangle diagrams for $(a)$ nucleon form factors in quark-diquark
model $(b)$ electron form factors in QED}
\label{fig.01}
\end{center}
\end{figure}

Although the calculation of the nucleon form factors based on a
quark-diquark model certainly differs from the calculation of the
electron form factors using quantum electrodynamics (QED), one may
still discern commonalities between the two apparently different
calculations.  For example, both calculations on the one-loop level
share essentially the same type of triangle diagram as shown in
Fig.~\ref{fig.01} for the computation of amplitudes.  While the contents
of the lines drawn in the two triangle diagrams are certainly different,
both calculations share the same type of one-loop integration for the
amplitudes given by three vertices connected by three propagators. In
particular, the structure of the two fermion lines intermediated
by a boson exchange is common in the two triangle diagrams and
may be generically identified as the four-fermion operator that
we discuss in this work.
Due to this commonality, it may be conceivable
to compute the two apparently different triangle amplitudes in a unified
way. Such a unified way of computation
is possible since the four-fermion operator can be Fierz-rearranged.

A similar idea of Fierz-rearranged four-fermion operators has been
developed in a rather different context of applications in the early
1980s. The basic idea of these developments was to provide a basis for the
one-boson-exchange interactions of baryons at low energy in the gluon
exchange which mediates quark-exchange scattering in conjunction with
quark interchange in a non-perturbative bag model framework~\cite{weber,
weber-maslow, bakker-et-al, bozoian-weber, beyer-weber}.  In elastic
nucleon-nucleon (NN) scattering, the four-fermion operator appears
from the gluon-exchange mediating quark-exchange scattering and becomes
bilocal when it is dressed with long-range quark-gluon correlations by
means of bag-model wavefunctions~\cite{weber}.  As this four-fermion
operator is Fierz-rearranged, the quark-interchange amplitude takes
on the usual local form for each nucleon that is expected from the
wealth of empirical knowledge at low energy~\cite{weber}.  The same
idea was applied to $\pi$N and $\pi \pi$ scattering as well as the
scattering involving hyperons~\cite{weber-maslow}.  A partial-wave
helicity-state analysis of elastic NN scattering was carried out in
momentum space~\cite{bakker-et-al} and a mesonic NN potential from an
effective quark interchange mechanism for non-overlapping nucleons was
obtained from the constituent quark model~\cite{beyer-weber}.  Also,
meson exchanges were introduced into the harmonic oscillator quark model
along with a lower component of the quark spinor~\cite{bozoian-weber}.

In this paper, we apply the Fierz-rearranged four-fermion operator
in the form factors shown in Fig.~\ref{fig.01} and present a global formula
to cover the triangle diagrams frequently used for the form factor calculations.  
The basic idea is presented in the next section, Sec.~\ref{sec:02},
and a simple illustration of this idea is given in Sec.~\ref{sec:02a} via
the self-energy calculation. In Sec.~\ref{sec:03}, we apply it to the form factor
calculations that involve triangle diagrams and present a corresponding
global formula.  Conclusion and outlook follow in Sec.~\ref{sec:04}.
Appendices A and B detail the four-fermion invariants in comparison with the 
well-known Fierz identities\cite{HJ87, Itzykson} and the manifestly covariant
calculation of form factors, respectively. 

\section{Basic Idea}
\label{sec:02}

\begin{figure}[htb]
\begin{center}
\includegraphics[width=80mm]{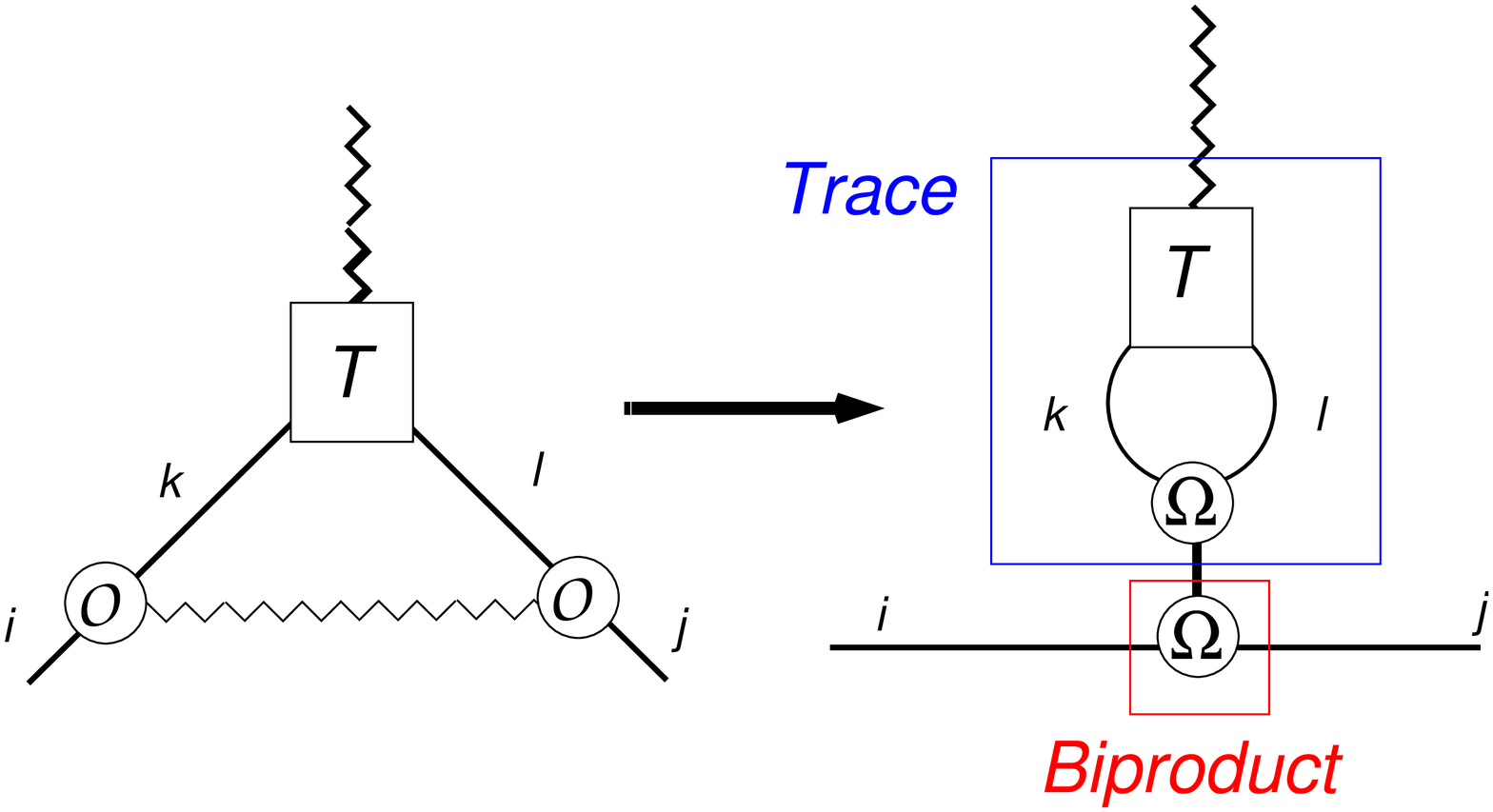}
\caption{Basic idea of Fierz-rearranged four-fermion operator in electromagnetic processes}
\label{fig.02}
\end{center}
\end{figure}

The basic idea of the four-fermion operator in electromagnetic
processes is depicted in Fig.~\ref{fig.02}, where a single-photon process
for a target nucleon is drawn as an illustration.  The left and right
portions of Fig.~\ref{fig.02} correspond to the amplitude intended for
computation and the equivalent amplitude after the four-fermion operator
is Fierz-rearranged, respectively.  In the left portion, a photon is
attached to the hadronic part which has the two intermediate quarks
denoted by the spinor indices $k$ and $\ell$ that inherit the fermion
number from the external nucleons, ${\bar u}_i$ and $u_j$, with the
corresponding spinor indices $i$ and $j$, respectively.  Here, the two
vertices, $O_{ik}$ and $O_{\ell j}$, connecting the nucleon and the
corresponding quark are linked to the scattering amplitude $T_{k \ell}$
where the photon interacts with the constituents from the target
nucleon: the rest of the constituents beside the quark is denoted by a
wiggly line below the corresponding quark and the loop integration over
the internal momentum is understood.  From this configuration of the
integrand in the amplitude, we may identify the four-fermion operator as
the multiplication of two vertices $O_{ik}O_{\ell j}$ and rearrange it as
\begin{eqnarray}
 \label{Fierz-rearrange}
 (O_\beta)_{ik}(O^\beta)_{\ell j} & = &
 {\mathsf{C}}^\beta_{\rm S} \delta_{ij}\delta_{\ell k} +
 {\mathsf{C}}^\beta_{\rm V} (\gamma_\mu)_{ij}(\gamma^\mu)_{\ell k}
 \nonumber \\
& + & {\mathsf{C}}^\beta_{\rm T}\left(\sigma_{\mu\nu}\right)_{ij}
 \left(\sigma^{\mu\nu}\right)_{\ell k}
\nonumber \\
 & + & {\mathsf{C}}^\beta_{\rm A} (\gamma_\mu \gamma_5)_{ij}(\gamma^\mu
 \gamma_5)_{\ell k} +
 {\mathsf{C}}^\beta_{\rm P} (\gamma_5)_{ij}(\gamma_5)_{\ell k}
\nonumber \\
 & = & \sum_{\alpha} {\mathsf{C}}^\beta_{\alpha} ({\Omega}_\alpha)_{ij}
 ({\Omega}^\alpha)_{\ell k},
\end{eqnarray}
where the index $\beta$ specifies the nature of
the operator $O$ in the vertex, whether it is scalar ($\rm S$), pseudoscalar ($\rm P$), vector ($\rm V$), axial-vector 
($\rm A$) or tensor ($\rm T$). Similarly, the index $\alpha$ specifies the nature of the rearranged operator $\Omega$ 
such that  $\Omega_S = {\bf  I}, \Omega_P = \gamma_5, \Omega_V = \gamma_\mu,\Omega_A=\gamma_\mu \gamma_5, \Omega_T = \sigma_{\mu\nu}= \frac{i}{2}[\gamma_\mu,\gamma_\nu]$, where the Lorentz indices appear obviously
for ${\rm V, A}$ and {\rm T} as denoted by $\mu$ and/or $\nu$.   Although
there are only 6 independent tensor operators in the full Dirac algebra,
we prefer to sum over the full number of 12 $\sigma$  tensors in
Eq.~(\ref{Fierz-rearrange}) using  Einstein's summation convention.
Appendix A details the comparison among different conventions\cite{HJ87, Itzykson} 
regarding in particular the compensating factor $\frac{1}{2}$ for this double counting in tensor operators
as well as the location of $\gamma_5$ in the axial-vector operator whether it be 
$\gamma_\mu \gamma_5$ or $\gamma_5 \gamma_\mu$.  

The Fierz coefficients ${\mathsf{C}}^\beta_{\alpha} (\alpha,\beta = {\rm
S, P, V, A, T})$ depend on the nature of the vertices $(O_\beta)_{ik}$
and $(O^\beta)_{\ell j}$. The operator $O$ is defined the same way as $\Omega$ is defined: i.e.,
$O_S={\bf I},  O_V=\gamma_\mu, O_T= \sigma_{\mu\nu}, O_A=\gamma_\mu \gamma_5,
O_P= \gamma_5$.
With this definition of operators $O$ and $\Omega$, 
we shall use the Fierz coefficients ${\mathsf{C}}^\beta_{\alpha}$ in Table~\ref{t1}
for different couplings $O_\beta({\rm and}\;\Omega_\alpha)$ (see also
Eq.~(\ref{aF6a}) in our Appendix A)\footnote{
Although our definition of the axial-vector operator, i.e. $O_A = \gamma_\mu \gamma_5$, differs
from the corresponding operator $\gamma_5 \gamma_\mu$ used in Ref.~\cite{HJ87}, Table~\ref{t1} is  
identical to the Fierz coefficients given in the same reference \cite{HJ87} (see e.g. Eq.~(\ref{aF6a})) 
because the swap of $\gamma_5$ and $\gamma_\mu$ doesn't matter on the level of the
four-fermion operator $(O_A)_{ik} (O^A)_{\ell j}$. See more details in Appendix A.}

For example, one can take the following coefficients from Table~\ref{t1}: $({\mathsf C}_{\rm S}, {\mathsf C}_{\rm V}, 
{\mathsf C}_{\rm T}, {\mathsf C}_{\rm A}, 
{\mathsf C}_{\rm P})=(3, 0, -1/2, 0, 3)$ and 
$(-1, -1/2, 0, -1/2, 1)$ for tensor vertices
$(\sigma_{\mu\nu})_{ik}(\sigma^{\mu\nu})_{\ell j}$ and the
axial-vector vertices 
$(\gamma_\mu \gamma_5)_{ik}(\gamma^\mu\gamma_5)_{\ell j}$, respectively.

\begin{table}
\caption{Fierz transformation coefficients of Eq.~(\ref{Fierz-rearrange})~\cite{HJ87}.}\label{t1}
\begin{tabular}{c|rrrrr} \hline\hline
  & S & V & T & A & ~P \\
\hline
 S  & 1/4 & 1/4 & 1/8 & $-1/4$ & ~1/4 \\
\hline
 V  & 1 & $-1/2$ & 0 & $-1/2$ & $~-1$ \\
\hline
 T  & 3 & 0 & $-1/2$ & 0 & ~3 \\
\hline
 A  & $-1$ & $-1/2$ & 0 & $-1/2$ & ~$1$ \\
\hline
 P  & 1/4 & $-1/4$ & 1/8 & $1/4$ & ~1/4 \\
\hline\hline
\end{tabular}
\end{table}
With this Fierz-rearrangement, we may write the integrand of the amplitude
(omitting the index $\beta$ for simplicity) as follows:
\begin{eqnarray}
\label{split}
{\bar u}_i O_{ik} T_{k \ell} O_{\ell j} u_j&=&
\sum_\alpha {\mathsf{C}}_{\alpha}\left({\bar u}_i {\Omega}^\alpha_{ij} u_j\right)
\left({\Omega}^\alpha_{\ell k} T_{k \ell} \right) \nonumber \\
&=& \sum_\alpha \left({\bar u} {\Omega}^\alpha u\right) {\mathsf{C}}_{\alpha}
{\rm Tr}[{\Omega}^\alpha  T],
\end{eqnarray}
where the external nucleon current (or biproduct ${\bar u} {\Omega}^\alpha
u$) part is now factorized from the internal scattering part given by
the trace of the quark loop ($ {\rm Tr}[{\Omega}^\alpha  T]$) as depicted in
the right portion of Fig.~\ref{fig.02}. 
With this rearrangement of the
same amplitude, one may get the general structure of the target hadron's
current more immediately and factorize the details of the internal probing
mechanism just due to the relevant constituents for the current of the
target hadron. It provides an efficient and unified way to analyze the
general structure of the amplitudes sharing the commonality of the same
type of diagram for the process.

\section{Simplest Illustration}
\label{sec:02a}

\begin{figure}[htb]
\begin{center}
 \includegraphics[width=80mm]{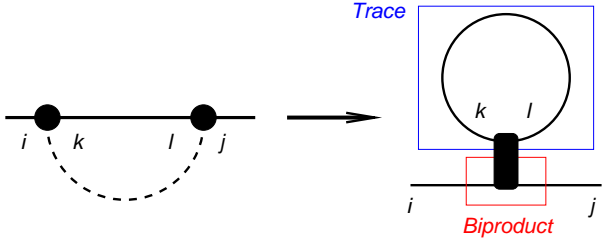}
\caption{Self-energy amplitude and the corresponding Fierz-rearrangement}
\label{fig.03}
\end{center}
\end{figure}

For an illustration of the basic idea, we start from the simple example
of a fermion self-energy amplitude which doesn't have any external photons but just has 
one loop due to an exchanged boson as shown in Fig. \ref{fig.03}.  
Such a process may occur in chiral perturbation theory to yield the self-energy of the nucleon due to the
surrounding pion cloud\cite{JTM}. Also, in the Yukawa model with a scalar coupling, the fermion self-energy 
due to a scalar boson has been investigated\cite{Nico-Ben-98}.

For the purpose of simple illustration, 
we consider here only scalar and pseudoscalar couplings (rather than 
the pseudovector coupling in chiral perturbation theory) and write the 
self-energy amplitude for a nucleon of mass $M$, four-momentum $p$ and spin $s$ 
in a unified formula both for scalar and pseudoscalar couplings:
\begin{eqnarray}
\label{SF}
\Sigma(p,s) &=& {\bar u}_i(p,s) \hat{\Sigma}_{ij} u_j(p,s)
\nonumber \\
&=& \Sigma_S {\bar u}(p,s) u(p,s) + \Sigma^\mu_V {\bar u}(p,s) \gamma_\mu u(p,s),
\end{eqnarray}
where modulo the appropriate normalization factor the self-energy operator
$\hat{\Sigma}_{ij}$ is given by
\begin{equation}
\label{SF-hat}
 {\hat \Sigma}_{ij} = \int \frac{d^4 k}{(2\pi)^4}
 \frac{O_{ik} (\slash{\!\!\!p}-\slash{\!\!\!k}+M)_{k\ell}O_{\ell j}}{D_k D_N}
\end{equation}
with $D_k = k^2 - m_X^2 + i \epsilon$ ($m_X$ is the intermediate meson mass)  and $D_N=(p-k)^2-M^2+
i \epsilon$. The four-fermion operator
$O_{ik} O_{\ell j}$ becomes ${\bf I}_{ik} {\bf I}_{\ell j} = \delta_{ik} \delta_{\ell j}$
for the scalar coupling theory, while it becomes $(\gamma_5)_{ik} (\gamma_5)_{\ell j}$
for the pseudoscalar coupling theory.
From Table~\ref{t1}, we get
\begin{eqnarray}
\label{scalar}
&&\hspace{-3em} \delta_{ik} \delta_{\ell j} =  \frac{1}{4} \delta_{ij} \delta_{\ell k} +  \frac{1}{4} (\gamma_\mu)_{ij} (\gamma^\mu)_{\ell k}
+ \frac{1}{8} (\sigma_{\mu\nu})_{ij} (\sigma^{\mu\nu})_{\ell k} \nonumber \\
&& - \frac{1}{4} (\gamma_\mu \gamma_5)_{ij}  (\gamma^\mu \gamma_5)_{\ell k}
+ \frac{1}{4} (\gamma_5)_{ij} (\gamma_5)_{\ell k} ,
\end{eqnarray}
and
\begin{eqnarray}
\label{pseudoscalar}
&&\hspace{-2em} (\gamma_5)_{ik} (\gamma_5)_{\ell j} =  \frac{1}{4} \delta_{ij} \delta_{\ell k} -  \frac{1}{4} (\gamma_\mu)_{ij} (\gamma^\mu)_{\ell k} 
+ \frac{1}{8} (\sigma_{\mu\nu})_{ij} (\sigma^{\mu\nu})_{\ell k} \nonumber\\ 
&&+ \frac{1}{4} (\gamma_\mu \gamma_5)_{ij}  (\gamma^\mu \gamma_5)_{\ell k}
+ \frac{1}{4} (\gamma_5)_{ij} (\gamma_5)_{\ell k}  ,
\end{eqnarray}
using Eq.~(\ref{Fierz-rearrange}) and Table~\ref{t1}. 
The structure given by 
$\sum_{\alpha} {\mathsf{C}}^\beta_{\alpha} ({\Omega}_\alpha)_{ij}
 ({\Omega}^\alpha)_{\ell k}$ in Eq.~(\ref{Fierz-rearrange}) is manifest both 
 in Eqs.~(\ref{scalar}) and (\ref{pseudoscalar}).
Now, using the Fierz-rearrangement given by Eq.~(\ref{Fierz-rearrange}), 
one may replace $O_{ik} O_{\ell j}$ with $\sum_{\alpha} {\mathsf{C}}^\beta_{\alpha} ({\Omega}_\alpha)_{ij}
 ({\Omega}^\alpha)_{\ell k}$.  Then, the multiplication of the operator factor $(\Omega^\alpha)_{\ell k}$ with
the factor $(\slash{\!\!\!p}-\slash{\!\!\!k}+M)_{k\ell}$ in Eq.~(\ref{SF-hat}) 
yields the trace ${\rm Tr}[{\Omega}^\alpha (\slash{\!\!\!p}-\slash{\!\!\!k}+M)]$
corresponding to the fermion loop shown in the right side of Fig.~\ref{fig.03}. 

Computing the trace ${\rm Tr}[{\Omega}^\alpha (\slash{\!\!\!p}-\slash{\!\!\!k}+M)]$, 
one can easily see that only $\alpha = {\rm S}$ and ${\rm V}$ survive while 
$\alpha = {\rm P, A,}$ and ${\rm T}$ vanish as expected from the structure of 
the fermion self-energy given by Eq.~(\ref{SF}). 
Since ${\rm Tr}[\slash{\!\!\!p}-\slash{\!\!\!k}+M] =4M$ and
${\rm Tr}[\gamma^\mu (\slash{\!\!\!p}-\slash{\!\!\!k}+M)] =4(p-k)^\mu$, 
we get $\Sigma_S$ and $\Sigma^\mu_V$ in Eq.~(\ref{SF}) as
\begin{eqnarray}
\label{SF-results}
 \Sigma_S &=& 4 {\mathsf{C}}_{\rm S} M \int \frac{d^4 k}{(2\pi)^4}
 \frac{1}{D_k D_N} , 
\nonumber \\
 \Sigma^\mu_V &=& 4 {\mathsf{C}}_{\rm V} \int \frac{d^4 k}{(2\pi)^4}
 \frac{(p-k)^\mu}{D_k D_N},
\end{eqnarray}
where ${\mathsf{C}}_{\rm S} = 1/4(1/4)$ and 
${\mathsf{C}}_{\rm V}= 1/4(-1/4)$ for the scalar (pseudoscalar) coupling case from Table~\ref{t1}.
This shows that both scalar and pseudoscalar coupling theories share the same
expressions given by Eq.~(\ref{SF-results}). From this unified formula, 
one can rather easily find a relationship between the two results, one from the scalar coupling 
theory and the other from the pseudoscalar theory: i.e.,
\begin{equation}
\label{relation}
(\Sigma_S)^{\rm S} = (\Sigma_S)^{\rm P} \hspace{1em} {\rm and} \hspace{1em} 
(\Sigma_V^\mu)^{\rm S} = - (\Sigma_V^\mu)^{\rm P} .
\end{equation}
The usual dimensional regularization method
 can be applied to obtain explicit results for $\Sigma_S$ and $\Sigma^\mu_V$
 after the four-dimensional integration over the internal four-momentum $k^\mu$ in the fermion loop.
They are found to be identical to the previous results\cite{JTM, Nico-Ben-98} obtained by the direct calculation without using the Fierz-rearrangement. It is amusing to notice that the results   
for the scalar coupling theory\cite{Nico-Ben-98} and the pseudoscalar coupling theory\cite{JTM} 
indeed satisfy the relationship given by Eq.~(\ref{relation}). Using the Fierz-rearrangement,
we now understand explicitly how and why they are related to each other.

\section{Application to Form Factors}
\label{sec:03}

We now apply the idea of the four-fermion operator and Fierz rearrangement
to the form factors shown in Fig.~\ref{fig.01} and present the result
that covers both the nucleon form factors in a quark-diquark model
and the electron form factors in QED. Using the four-fermion method
illustrated in Sec.~\ref{sec:02}, the current operator $J^\mu$ in the amplitude
${\bar u}(p')J^\mu u(p)$ from the triangle diagram with the external
fermion mass $M$, the internal fermion mass $m$ and the intermediate
boson mass $m_X$ can be given by (modulo normalization)
\begin{equation}
\label{current}
 J^\mu = ig^2 \int \frac{d^4 k}{(2\pi)^4} \frac{ {\mathsf{N}}^\mu }
 {D_{p_1}D_{p_2}D_k},
\end{equation}
where $p_{1(2)}=p(p')-k$, $D_{p_i} = p_i^2-m^2+i\varepsilon$,
$D_k  =  k^2 - m_X^2+i\varepsilon$ and ${\mathsf{N}}^\mu$ is the numerator
of the amplitude corresponding to the triangle diagram (e.g. Fig.~\ref{fig.01}). 
Using the Fierz rearrangement given by Eq.~(\ref{Fierz-rearrange}), the numerator
${\mathsf{N}}^\mu$ can be written as 
\begin{equation}
\label{4F7-1}
 {\mathsf{N}}^\mu
  =  \sum_\alpha {\mathsf{C}}_{\alpha}{\rm Tr}
 \left[\left(\slash \!\!\!p_2+m \right)\gamma^\mu
 \left(\slash \!\!\!p_1+m \right) \Omega^\alpha\right] \Omega_\alpha,
\end{equation}
where $\alpha={\rm S,V,T,A,P}$ and the corresponding 
$\Omega^\alpha =  {\mathbf I},\gamma^\nu,\sigma^{\nu\delta},
\gamma^\nu\gamma_5, \gamma_5$ with the dummy Lorentz indices
$\nu$ and $\delta$. As expected from the parity conservation
in the electromagnetic current, $\alpha = {\rm P}$ never contributes
to ${\mathsf{N}}^\mu$: i.e., ${\rm Tr}
 \left[\left(\slash \!\!\!p_2+m \right)\gamma^\mu
 \left(\slash \!\!\!p_1+m \right) \gamma_5 \right]= 0$.
Thus, after the trace calculation, we get 
\begin{eqnarray}
{\mathsf N}^\mu
& = & 4 \left[ {\mathsf{C}}_{\rm S} \hspace{.2em} m\left(p_1 + p_2\right)^\mu \rule{0mm}{4mm}  \right. 
 \nonumber \\
 &+& 
 \left.
 {\mathsf{C}}_{\rm V}  \left \{(m^2 -p_1\cdot p_2)\gamma^\mu 
+  p_2^\mu (p_1\cdot\gamma) +(p_2\cdot\gamma)p_1^\mu \right \}
\right.
\nonumber \\
 & + & \left.  i  {\mathsf{C}}_{\rm T} \hspace{.2em} m
 \left(g^{\mu \alpha}q^\beta - g^{\mu \beta} q^\alpha \right)
 \sigma_{\alpha \beta}
 \right. 
\nonumber \\
 &-& 
 \left.
  i {\mathsf{C}}_{\rm A}\epsilon^{\mu \nu \alpha \beta}
 \left(p_2\right)_\nu \gamma_\al\left(p_1\right)_\beta\gamma_5
\right ],
\end{eqnarray}
where $q=p'-p$. 

Now, using the usual Feynman parametrization for the loop integration, 
the denominator of the integrand in Eq.~(\ref{current}) yields
\begin{eqnarray}
\label{4F8}
&&\hspace{-1em} \frac{1}{D_{p_1}D_{p_2}D_k}   \nonumber \\
&&= 2\int_0^1 dx \int_0^{1-x} \frac{dy}{\left[D_k+x\left(D_{p_1}-D_k\right)
 +y\left(D_{p_2}-D_k\right) \right]^3}
\nonumber \\
 && =2\int_0^1 dx \int_0^{1-x} \frac{dy}{(k^{\prime 2}-M_{\rm cov}^2)^3},
\end{eqnarray}
where in the second line we used the shifted momentum $k^\prime = k -
xp - y p^\prime$ and defined
\begin{eqnarray}
\label{denominator}
M_{\rm cov}^2 &=& (x+y)m^2 + (1-x-y)m_X^2 -xy q^2 
\nonumber \\
&-& (x+y)(1-x-y)M^2
\end{eqnarray}
with the on-shell condition $p^2=p^{\prime\, 2} = M^2$.
Then, the current operator $J^\mu$ given by Eq.~(\ref{current}) becomes
\begin{equation}
\label{4F10}
 J^\mu = 2 i g^2\int_0^1 dx \int_0^{1-x} dy
 \int \frac{d^4 k^{\prime}}{(2\pi)^4}
 \frac{\widetilde{{\mathsf{N}}}^\mu}{\left(k^{\prime 2}-M_{\rm cov}^2\right)^3},
\end{equation}
where the numerator with shifted momentum is now given by
\begin{eqnarray}
\label{4F11}
 \widetilde{{\mathsf{N}}}^\mu & = & 4 \left[ {\mathsf{C}}_{\rm S} (1-x-y)m(p^\prime+p)^\mu
 \rule{0mm}{5mm} \right.
\nonumber \\
 & + & {\mathsf{C}}_{\rm V}\left \{ \rule{0mm}{5mm} \left(m^2 - k^{\prime 2}-(1-x-y)^2M^2
\rule{0mm}{5mm}  \right.\right.
 \\
 &&\quad +
 \left.\left.
 (1-x-y+2xy)\frac{q^2}{2}\right)\gamma^\mu \right.
\nonumber \\
 && \quad + \left. 2k^{\prime \mu}\slash\!\!\!k^\prime +\frac{(1-x-y)^2}{2}
 (p^\prime+p)^\mu (\slash\!\!\!p+\slash\!\!\!p') \right. 
 \nonumber \\
 && \quad - \left. \frac{1-(x-y)^2}{2}q^\mu \slash\!\!\!q \right \}
\nonumber \\
& + &\left.  2 i  {\mathsf{C}}_{\rm T} \hspace{.2em} m \hspace{.2em} g^{\mu \alpha} q^\beta
 \sigma_{\alpha \beta} \right. \nonumber \\ 
 & + & 
 \left.
 i{\mathsf{C}}_{\rm A}(1-x-y)\epsilon^{\mu \nu \alpha \beta}
 p_\nu \gamma_\al p^{\prime}_{\beta}\gamma_5
 \rule{0mm}{4mm} \right ].\nonumber
\end{eqnarray}

Although one expects to get $J^\mu=\gamma^\mu F_1(q^2)
+i \frac{\sigma_{\mu\nu}q^\nu}{2M}F_2(q^2)$, our result for $J^\mu$
appears to exhibit not only the vector and tensor currents but also the
scalar and axial vector currents. This issue can be resolved by the Gordon
decomposition and a similar extension namely
\begin{eqnarray}
\label{GD}
 \left(p^\prime+p\right)^\mu & \longrightarrow &
 2M\gamma^\mu -i\sigma^{\mu \nu} q_\nu,
\nonumber \\
 i\epsilon^{\mu \nu \alpha \beta}\gamma_5 p_\nu\gamma_\alpha p^\prime_\beta
 & \longrightarrow & \frac{q^2}{2}\gamma^\mu-iM\sigma^{\mu \nu}q_\nu.
\end{eqnarray}
Using Eq.~(\ref{GD}), we get the expected decomposition of $J^\mu$ in terms of just vector
and tensor currents and find the form factors  ($i=1,2$) as follows:
\begin{equation}
\label{4F17-1}
 F_i\left(q^2\right) = 8 i g^2 \int_0^1dx\int_0^{1-x}dy
 \int d^4 k^\prime \frac{\mathsf{\tilde N}_i}{\left({k^\prime}^2
 - M_{\rm cov}^2\right)^3}\,,
\end{equation}
where
\begin{eqnarray}
\label{4F18-1}
 \mathsf{\tilde N}_1 & & \hspace{-1em}=  2mM(1\!-\!x\!-\!y){\mathsf C}_{\rm S} 
+ (1\!-\!x\!-\!y)\frac{q^2}{2}{\mathsf C}_{\rm A}
\nonumber\\
&&\hspace{-2em} 
+ \left[m^2+(1\!-\!x\!-\!y)^2M^2+(1\!-\!x\!-\!y\!+\!2xy)\frac{q^2}{2}
 -\frac{k^2}{2}\right]{\mathsf C}_{\rm V},
\nonumber\\
 \mathsf{\tilde N}_2 && \hspace{-1em}= -2mM(1\!-\!x\!-\!y){\mathsf C}_{\rm S}
 - 2(1\!-\!x\!-\!y)M^2{\mathsf C}_{\rm A}
 \nonumber\\
 &&-2(1\!-\!x\!-\!y)^2M^2 {\mathsf C}_{\rm V}
 + 4 m M{\mathsf C}_{\rm T} .
\end{eqnarray}
Apparently, $F_1$ is UV divergent and requires a regularization
along with the renormalization set by the normalization condition
$F_1(0)=1$. More explicit expressions for $F_1(q^2)$ and $F_2(q^2)$ are derived
in Appendix \ref{sec:Yukawa} using dimensional regularization 
with Wick rotation.  

We may point out that the results in Eq.~(\ref{4F17-1}) can cover 
not only the nucleon form factors in a quark-diquark model, whether the diquark is
scalar or axial-vector, but also the electron form factors in QED
taking the corresponding Fierz coefficients and masses. For example,
from Table \ref{t1}, 
${\mathsf C}_{\rm S}={\mathsf C}_{\rm V}=2{\mathsf C}_{\rm T}= -{\mathsf
C}_{\rm A}={\mathsf C}_{\rm P}=1/4$ if the diquark is taken as a scalar boson
, while ${\mathsf C}_{\rm S}=-1,~{\mathsf
C}_{\rm V}=-1/2,~{\mathsf C}_{\rm T}=0,~{\mathsf C}_{\rm A}=-1/2$, and
${\mathsf C}_{\rm P}=1$  if the diquark is taken as an axial vector boson in a 
quark-diquark model for the nucleon form factors.  
For the electron form factors in QED, one should take 
of course ${\mathsf C}_{\rm S}=1,~{\mathsf
C}_{\rm V}=-1/2,~{\mathsf C}_{\rm T}=0,~{\mathsf C}_{\rm A}=-1/2$, and
${\mathsf C}_{\rm P}=-1$. Also,
$M, m$ and $m_X$ are the nucleon, quark and diquark masses for the nucleon form
factors, while $M=m$ is the electron mass in the QED calculation of the
electron form factors. 
It is interesting to see that the Fierz coefficient ${\mathsf C}_{\rm P}$  
appears neither in the nucleon form factors nor in the electron form factors 
reflecting the parity conservation both in the strong and electromagnetic interactions.
We should note, however, that the disappearance of
${\mathsf C}_{\rm P}$ in Eq.~(\ref{4F17-1}) is not coming from the Fierz
rearrangement itself but coming from the trace calculation reflecting
the conservation of parity in the single photon process: e.g. a pion
can never decay into a single photon but can decay into two photons.
Thus, we may expect that the contribution from $\Omega_{\rm P}$ would 
show up in the amplitude defined in a process involving two photons 
such as the generalized parton distributions in deeply virtual Compton scattering.

Finally, we note that the usual decomposition of $J^\mu=\gamma^\mu F_1(q^2)
+i \frac{\sigma_{\mu\nu}q^\nu}{2M}F_2(q^2)$ in terms
of vector and tensor currents with the Dirac ($F_1$) and Pauli ($F_2$)
form factors is just one of six possible decompositions:
\begin{eqnarray}
\label{equiv-FF}
 J^\mu &=& \gamma^\mu F_1 + i \frac{\sigma^{\mu\nu}q_\nu}{2 M} F_2
\nonumber \\
 &=& \gamma^\mu (F_1+F_2) + \frac{(p+p')^\mu}{2 M} F_2
\nonumber \\
 &=& \frac{(p+p')^\mu}{2 M}\frac{4M^2 F_1+q^2 F_2}{4M^2-q^2} 
  \nonumber \\
&-&
 i\epsilon^{\mu\nu\alpha\beta}
 \gamma_5\gamma_\nu p_\alpha p'_\beta \frac{2(F_1+F_2)}{4M^2-q^2}
\\
 &=& \frac{(p+p')^\mu}{2 M} F_1 
 + i \frac{\sigma^{\mu\nu}q_\nu}{2 M} (F_1+F_2)
\nonumber \\
 &=& \gamma^\mu (F_1 + \frac{q^2}{4M^2}F_2)-i\epsilon^{\mu\nu\alpha\beta}
 \gamma_5\gamma_\nu p_\alpha p'_\beta \frac{F_2}{2M^2}
\nonumber \\
 &=&  i \frac{\sigma^{\mu\nu}q_\nu}{2 M}(\frac{4M^2}{q^2}F_1+ F_2)+
 i\epsilon^{\mu\nu\alpha\beta}
 \gamma_5\gamma_\nu p_\alpha p'_\beta \frac{2F_1}{q^2} .
 \nonumber
\end{eqnarray}
One should note, however, that the equivalence presented in Eq.~(\ref{equiv-FF})
meant the equality on the level of matrix elements, e.g. ${\bar u} J^\mu u$, but
not on the level of operators themselves. In other words, Eq.~(\ref{equiv-FF}) is valid only for
the spin-1/2 fermion case such as the nucleon. Thus, for the nucleon target,  
these six different decompositions in Eq.~(\ref{equiv-FF}) are all
equivalent. Any particular choice of decomposition may depend on a 
matter of convenience and/or effectiveness in the given situation
of computation.

\section{Conclusion}
\label{sec:04}
The idea of rearranging four-fermion operators provides an effective way
to analyze hadronic processes. It factorizes the details of the internal
probing mechanism from the external global structure owing to the target
hadrons. In this work, we illustrated the idea of Fierz-rearrangement to 
the fermion self-energy and electromagnetic form factor calculations.
Processes sharing a certain commonality (e.g. the same type
of diagrams) may be described in a unified way. 
For instance, whether the mesons surrounding the nucleon are scalar or 
pseudoscalar bosons, the Fierz rearrangement of 
the four-fermion operators can be used to yield a unified expression for the
nucleon self-energy amplitude and provide a relationship between the 
two amplitudes, one for the scalar coupling and the other for the pseudoscalar
coupling. 
Likewise, the electromagnetic nucleon form factors in a quark-diquark model
and the electron form factors in QED can be given by a unified expression 
based on the commonality of sharing the same type of  diagram, e.g. 
the triangle diagrams shown in Fig. \ref{fig.01}.   
Moreover, the quark-diquark calculations of baryon form factors using the idea of 
rearranging four-fermion operators proposed in this work may provide a unified expression
which can cover all types of diquarks such as scalar, pseudoscalar, vector, axial-vector and 
tensor diquarks. With this idea, we can offer a clear understanding of the interrelationships among 
different calculations sharing a commonality. 

While we presented only the basic idea and a few simple examples in this paper,
we may foresee a great potential for further application to other hadronic processes.  
In particular, the application to the two-photon processes would
be interesting since the generalized hadronic tensor structure of DVCS
still needs further investigation\cite{GPD-Review} in view of forthcoming experiments
with the 12 GeV upgrade at JLab. Work along this line is in progress.

\acknowledgments
This work is supported by the US Department of Energy
(No. DE-FG02-03ER41260). HMC and ATS acknowledge partial supports
from KRF Grant (KRF-2010-0009019) and CNPq (Processo no. 201.902/2010-9),
respectively.

\appendix
\section{Conventions in Fierz identities for Dirac matrices}
\label{App.A}

A Fierz identity is an identity that allows one to rewrite bilinears of
the product of two spinors as a linear combination of products of the
bilinears of the individual spinors. In all, 16 bilinear terms can be
constructed using bispinors ${\bar a}$ and $b$. The linear combinations
of these terms form five different types of Lorentz-covariant quantities,
${\bar a} b, {\bar a} \gamma^\mu b, {\bar a} \sigma^{\mu\nu} b, {\bar
a} \gamma^5\gamma^\mu b, {\bar a} \gamma^5 b$, where $\sigma^{\mu\nu}=
\frac{i}{2}[\gamma^\mu,\gamma^\nu]$. These covariants are normalized
as follows
\begin{eqnarray}
&& {\bf I}\cdot {\bf I}=1,\; \gamma_\mu\gamma^\mu=4,\;
 \sigma_{\mu\nu}\sigma^{\mu\nu}=12,\nonumber \\
&&(\gamma_5\gamma_\mu)(\gamma^5\gamma^\mu)=-4,\;
 \gamma_5\gamma^5=1.
\label{aF1}
\end{eqnarray}
There are several ways of constructing the Lorentz scalar out of four
bispinors ${\bar a},b,{\bar c}, d$~\cite{HJ87,Itzykson,Okun}. According 
to the convention by Weber~\cite{HJ87}, the five Lorentz scalars can be 
constructed out of four bispinors ${\bar u}_1, u_2,{\bar u}_3, u_4$ as follows:
\begin{eqnarray}
\label{aF3}
 {\rm S-variant}&:& S^{W}(4,2;3,1) =  ({\bar u}_4 u_2)({\bar u}_3 d_1),
\nonumber\\
 {\rm V-variant}&:& V^{W}(4,2;3,1) = ({\bar u}_4 \gamma^\mu u_2)({\bar u}_3 \gamma_\mu u_1),
\nonumber\\
{\rm T-variant}&:& T^{W}(4,2;3,1) = ({\bar u}_4 \sigma^{\mu\nu} u_2)({\bar u}_3 \sigma_{\mu\nu} u_1),
\nonumber\\
 {\rm A-variant}&:& A^{W}(4,2;3,1) =({\bar u}_4 \gamma^5\gamma^\mu u_2)({\bar u}_3 \gamma_5\gamma_\mu u_1),
\nonumber\\
 {\rm P-variant}&:& P^{W}(4,2;3,1) =({\bar u}_4 \gamma^5 u_2)({\bar u}_3 \gamma_5 u_1).
\end{eqnarray}
Counting only the independent tensors, $\sigma^{\mu\nu}$
with $\mu <\nu$, the sixteen matrices $({\bf I},
\gamma^\mu,\sigma^{\mu\nu},\gamma^5\gamma^\mu,\gamma_5)$ form a complete
set so that any one of the above variants can be expressed as a
linear combination of variants with a changed sequence of spinors:
\begin{equation}
\label{aF4}
 ({\bar u}_4 {W}^i u_2)({\bar u}_3 {W}_i u_1) = \sum_k \mathsf{C}^{i}_{k}({\bar u}_4 {W}^k u_1)({\bar u}_3 {W}_k  u_2),
\end{equation}
where
\begin{equation}
\label{aF5}
 W_S ={\bf I},\;\; W_V =\gamma_\mu,\;\; W_T =\sigma_{\mu\nu},\;\;
 W_A =\gamma_5\gamma_\mu,\;\; W_P =\gamma_5.
\end{equation}
Our vertex operators denoted by $O_\beta$ as well as the rearranged operators denoted by $\Omega_\alpha$
are the same as Weber's operators $W_i$ except for the swap of $\gamma_5$ and $\gamma_\mu$ in the
axial vector operator: i.e. $O_S=W_S,  O_V=W_V, O_T= W_T, O_A=-W_A,
O_P= W_P$. Thus, the coefficients $\mathsf{C}^{i}_{k}$ in Eq.~(\ref{aF4}) are identical to the Fierz coefficients
given by Table \ref{t1}. More explicitly, one may write Eq.~(\ref{aF4}) as a matrix equation: i.e.,
\begin{eqnarray} \label{aF6a}
&&\hspace{-2em} \left[\begin{array}{c}
 S^W \\
 V^W \\
 T^W \\
 A^W \\
 P^W
 \end{array}
\right](4,2;3,1) = 
\\
&& \quad\quad \frac{1}{4}
\left[
\begin{array}{rrrrr}
 1 & 1 & \frac{1}{2} & -1 & 1 \\
 4 & -2 & 0 & -2 & -4 \\
 12 & 0 & -2 & 0 & 12  \\
 -4 & -2 & 0 & -2 & 4 \\
 1 & -1 & \frac{1}{2} & 1 & 1 \\
\end{array}
\right]
\left[\begin{array}{c}
 S^W \\
 V^W \\
 T^W \\
 A^W \\
 P^W
 \end{array}
 \right](4,1;3,2). \nonumber 
\end{eqnarray} 


\noindent
For example, either from Table~\ref{t1} or Eq.~(\ref{aF6a}), one can read off  
the following
coefficients: $(\mathsf{C}^{\rm T}_{\rm S} =3, \mathsf{C}^{\rm T}_{\rm
V}=0, \mathsf{C}^{\rm T}_{\rm T}=-1/2,  \mathsf{C}^{\rm T}_{\rm A}=0,
\mathsf{C}^{\rm T}_{\rm P}=3)$ for the tensor product or T-variant $({\bar u}_4 \sigma^{\mu\nu}
u_2)({\bar u}_3 \sigma_{\mu\nu} u_1)$ and $(\mathsf{C}^{\rm A}_{\rm
S}=-1, \mathsf{C}^{\rm A}_{\rm V}=-1/2, \mathsf{C}^{\rm A}_{\rm T}=0,
\mathsf{C}^{\rm A}_{\rm A}=-1/2, \mathsf{C}^{\rm A}_{\rm P}=1)$
for the axial-vector product or A-variant $({\bar u}_4 \gamma^5\gamma^\mu u_2)({\bar u}_3
\gamma_5\gamma_\mu u_1)$.

On the other hand, according to the convention by Itzykson and Zuber
(IZ)~\cite{Itzykson}, the five different Lorentz-covariant quantities
are taken as 
$\{{\bar a} b,~{\bar a} \gamma^\mu b,~{\bar a}\sigma^{\mu\nu} b, {\bar a} \gamma^5\gamma^\mu b, {\bar a}(i\gamma^5) b\}$. Note here that
the factor $i$ in front of $\gamma^5$ makes the pseudoscalar operator $i \gamma^5$
hermitian. These five Lorentz-covariant quantities are paired with their partners
$\{{\bar a} b,~{\bar a} \gamma_\mu b,~{\bar a}\sigma_{\mu\nu} b, 
{\bar a}~\gamma_\mu \gamma_5b,~{\bar a}(-i\gamma_5) b\}$ to construct
the corresponding five Lorentz scalars. Using the four
bispinors ${\bar u}_1, u_2,{\bar u}_3, u_4$, we may write those Lorentz scalars as follows:
\begin{eqnarray}
\label{aF6}
  S^{IZ}(4,2;3,1) &=&  ({\bar u}_4 u_2)({\bar u}_3 d_1),
 \nonumber\\
  V^{IZ}(4,2;3,1) &=& ({\bar u}_4 \gamma^\mu u_2)({\bar u}_3 \gamma_\mu u_1),
 \nonumber\\
 T^{IZ}(4,2;3,1) &=& \frac{1}{2}({\bar u}_4 \sigma^{\mu\nu} u_2)({\bar u}_3 \sigma_{\mu\nu} u_1),
 \nonumber\\
  A^{IZ}(4,2;3,1) &=& ({\bar u}_4 \gamma^5\gamma^\mu u_2)({\bar u}_3 \gamma_\mu\gamma_5 u_1),
 \nonumber\\
  P^{IZ}(4,2;3,1) &=& ({\bar u}_4 \gamma^5 u_2)({\bar u}_3 \gamma_5 u_1).
\end{eqnarray}
Note here that the factor $i$ in $i \gamma^5$ and $-i$ in $-i \gamma_5$ are
not written explicitly in $P^{IZ}$, because they cancel out. Also,  
the usual summation convention is used in $T^{IZ}$ to sum over all 
twelve tensor operators. Because only six independent tensor operators exist, 
the factor of $\frac{1}{2}$ is introduced in $T^{IZ}$ to compensate for 
this double counting. Finally, we note that the $T$- and $A$-variants defined in Eq.~(\ref{aF6}) are
different from those in Eq.~(\ref{aF3}), i. e., $T^{IZ}=\frac{1}{2} T^{W}$
and $A^{IZ}=-A^{W}$.  Accordingly, the coefficients $\mathsf{C}^{i}_{k}$
in the IZ convention of Lorentz scalars are given by
\begin{eqnarray} 
\label{aF7}
&&\hspace{-2em} \left[\begin{array}{c}
 S^{IZ} \\
 V^{IZ} \\
 T^{IZ} \\
 A^{IZ} \\
 P^{IZ}
 \end{array}
\right](4,2;3,1) =
\\
&& \quad\quad \frac{1}{4}
\left[
\begin{array}{rrrrr}
 1 & 1 & 1 & 1 & 1 \\
 4 & -2 & 0 & 2 & -4 \\
 6 & 0 & -2 & 0 & 6  \\
 4 & 2 & 0 & -2 & -4 \\
 1 & -1 & 1 & -1 & 1 \\
\end{array}
\right]
\left[\begin{array}{c}
 S^{IZ} \\
 V^{IZ} \\
 T^{IZ} \\
 A^{IZ} \\
 P^{IZ}
 \end{array}
 \right](4,1;3,2).\nonumber
\end{eqnarray}

\section{Explicit Results of Form Factors in Eq.~(\ref{4F17-1})}
\label{sec:Yukawa}

Using the four-fermion method
illustrated in Sec.~\ref{sec:02} and the usual Feynman parametrization
for the loop integration, we computed the triangle diagrams shown in Fig.~\ref{fig.01}
for the electromagnetic form factors of the spin 1/2 target particle
and obtained Eq.~(\ref{4F17-1}) as presented in Sec.~\ref{sec:03}.
Since the momentum integral in Eq.~(\ref{4F17-1}) diverges for the
$k^{\prime 2}$ and $k^{\prime \mu}k^{\prime \alpha}$ terms in the
ultraviolet region, we need to regularize it. 
In this Appendix, we perform the four-dimensional $k^\prime$ integration
using the Wick rotation to Euclidean space, 
$k^{\prime} = i\kappa_{\rm E}$, and the dimensional regularization 
to find the more explicit expressions for $F_1(q^2)$ and $F_2(q^2)$.  
Then, Eq.~(\ref{4F10}) is rewritten as
\begin{equation}
\label{4F12}
 J^\mu  = 2 \mu^{4-d}g^2\int_0^1 dx \int_0^{1-x} dy
 \int \frac{d^d \kappa_{\rm E}}{(2\pi)^d}
 \frac{\widetilde{{\mathsf{N}}}^\mu_{\rm E}}{\left(\kappa^{2}_{\rm E}
 +M_{\rm cov}^2\right)^3},
\end{equation}
where $\mu^{d-4}$ is the usual mass factor that comes in to compensate
the change in the dimensionality of the momentum integration.
Using the property of symmetric integration for
$\kappa^{\mu}_{\rm E}\kappa^{\alpha}_{\rm E}=\kappa^2_{\rm E}g^{\mu \alpha}/d$,
we have now the following numerator which corresponds to Eq.~(\ref{4F11}):
\begin{eqnarray}
\label{4F14}
 \widetilde{{\mathsf{N}}}^\mu_{\rm E} & = & d \left[ {\mathsf{C}}_{\rm S}
(1-x-y)m(p^\prime+p)^\mu \rule{0mm}{4mm} \right.
\nonumber \\
 & + &{\mathsf{C}}_{\rm V} \left \{ \rule{0mm}{4mm} \left(m^2 \!+\! \left(1-\frac{2}{d}\right)\kappa^{2}_{\rm E}
 -(1-x-y)^2M^2 \nonumber
 \right.\right.
 \\
&& \quad +
\left.\left.
(1-x-y+2xy)\frac{q^2}{2}\right)\gamma^\mu \right.
 \\
 && \quad + \left. (1-x-y)^2\,M\,(p^\prime+p)^\mu  \rule{0mm}{5mm}  \right \}
\nonumber \\
& + & \left. 2 i {\mathsf{C}}_{\rm T} \hspace{.2em} m \hspace{.2em} 
\sigma^{\mu \nu} q_\nu \right. \nonumber \\
& + &
\left.
i{\mathsf{C}}_{\rm A}(1-x-y)\epsilon^{\mu \nu \alpha \beta}
 p_\nu \gamma_\al \gamma_5 p^{\prime}_{\beta}
 \rule{0mm}{4mm} \right ].\nonumber
\end{eqnarray}

Now, using Eq.~(\ref{GD}), we can decompose $J^\mu$
in terms of vector and tensor currents and find the form factors
$F_i(i=1,2)$:
\begin{equation}
\label{4F17}
 F_i\left(q^2\right) = 2 d\mu^{4-d}g^2\int_0^1dx\int_0^{1-x}dy \int
 \frac{d^d\kappa_{\rm E}}{(2\pi)^d}\,\frac{\mathsf{N}_i}{\left(\kappa^2_{\rm E}
 +M_{\rm cov}^2\right)^3}\,,
\end{equation}
where
\begin{eqnarray}
\label{4F18}
 \mathsf{N}_1 & = & 2mM(1\!-\!x\!-\!y){\mathsf C}_{\rm S} +
(1\!-\!x\!-\!y)\frac{q^2}{2}{\mathsf C}_{\rm A}
\nonumber\\
 && +\left[m^2+(1\!-\!x\!-\!y)^2M^2+(1\!-\!x\!-\!y\!+\!2xy)\frac{q^2}{2}
 \right.
 \nonumber \\
 &&\quad\quad +
 \left.
 \left(1\!-\!\frac{2}{d}\right)\kappa_{\rm E}^2\right]{\mathsf C}_{\rm V},
\nonumber \\
 \mathsf{N}_2 & = & -2mM(1\!-\!x\!-\!y){\mathsf C}_{\rm S}-2(1\!-\!x\!-\!y)^2M^2
 {\mathsf C}_{\rm V}
 \nonumber \\
&& - 2(1\!-\!x\!-\!y)M^2{\mathsf C}_{\rm A} + 4 m M{\mathsf C}_{\rm T}.
\end{eqnarray}

The momentum integration can be performed in Eq.~(\ref{4F17}) using the
following standard results:
\begin{eqnarray}
\label{4F24}
d\mu^{4-d}\int d^d\kappa_{\rm E} \frac{1}{\left(\kappa_{\rm E}^2+M_{\rm cov}^2\right)^3} & = & \frac{2\pi^2}{M_{\rm cov}^2}+ {\mathcal O}(\epsilon),
\nonumber \\
d\mu^{4-d}\int d^d\kappa_{\rm E} \frac{\kappa_{\rm E}^2}{\left(\kappa_{\rm E}^2+M_{\rm cov}^2\right)^3} & = &
 \\
&& \hspace{-3em}
 \pi^2\left(2-\epsilon\right)^2\left(\frac{\mu^2}{\pi M_{\rm cov}^2}\right)^\epsilon \Gamma(\epsilon),\nonumber 
\end{eqnarray}
where on the right hand side we used the definition $2\epsilon = 4 - d$.
Expanding  the second result above for small $\epsilon$, we have
\begin{eqnarray}
\label{4F25}
 \left(1-\frac{2}{d}\right)d\mu^{4-d}\int d^d\kappa_{\rm E}
 \frac{\kappa_{\rm E}^2}{\left(\kappa_{\rm E}^2+M_{\rm cov}^2\right)^3} &=&
 \\
&& 
\hspace{-12em}
2\pi^2\left\{\frac{1}{\epsilon}-\gamma -\frac{3}{2}
 + \ln\left(\frac{\mu^2}{\pi M_{\rm cov}^2}\right)+
 {\mathcal O}(\epsilon)\right\}. \nonumber
\end{eqnarray}
We finally get
\begin{eqnarray}
\label{4F26}
&& \hspace{-1.3em} F_1(q^2) = \nonumber \\
&& \hspace{-.5em}\frac{g^2}{4\pi^2} \int_0^1dx\int_0^{1-x} dy
 \biggl\{
 \left[
 \frac{1}{\epsilon}-\gamma-\frac{3}{2}+\ln(\frac{\mu^2}{\pi M^2_{\rm cov}})
 \right]{\mathsf C}_V \biggr.
\nonumber \\
&&\hspace{-.5em}+ 
\biggl.
\frac{2Mm (1-x-y ){\mathsf C}_{\rm S} +
 (1-x-y)\frac{q^2}{2}{\mathsf C}_A}{M^2_{\rm cov}}
\\
&& \hspace{-.5em}+
\frac{\left[m^2+ (1-x-y)^2M^2+(1-x-y+2xy)\frac{q^2}{2}\right]{\mathsf C}_{\rm V}}{M_{\rm cov}^2}
 \biggr\}, \nonumber
\end{eqnarray}
and
\begin{eqnarray}
\label{4F27}
&& \hspace{-2em} F_2(q^2)  = \nonumber \\
&& \hspace{-.5em} \frac{g^2}{2\pi^2} \int_0^1 dx \int_0^{1-x} dy
 \biggl\{\frac{2Mm{\mathsf C}_{\rm T}-Mm\left(1-x-y\right){\mathsf C}_{\rm S}}
 {M^2_{\rm cov}} \biggr.
 \nonumber \\
&& \hspace{-.5em} - 
\biggl.
\frac{\left(1-x-y\right)^2M^2{\mathsf C}_{\rm V}
 +\left(1-x-y\right)M^2{\mathsf C}_{\rm A}}{M^2_{\rm cov}}\biggr\}.
\end{eqnarray}

We now check whether the form factors given by Eqs.~(\ref{4F26})
and~(\ref{4F27}) are consistent with specific spectator particles,
such as scalar meson exchange, vector meson/photon exchange, 
etc. To do that, we need to consider how the different coefficients
are expressed, and this is achieved by using appropriate Fierz
rearrangements in  $O_{ik}O_{\ell j}$ given by
Eq.~(\ref{Fierz-rearrange}) and Table~\ref{t1}.
For
example, if we want the coefficients for the scalar meson exchange
(e.g. the Yukawa model), the proper coefficients are
$\mathsf{C}_{\rm S} =\mathsf{C}_{\rm V}=2 \mathsf{C}_{\rm T}=
- \mathsf{C}_{\rm A}=\mathsf{C}_{\rm P}=\frac{1}{4}$.
Substituting these values in Eqs.~(\ref{4F26}) and~(\ref{4F27}) we get:
\begin{eqnarray}
\label{Yu1}
&&\hspace{-2.5em} F_1^{\mathsf{scalar}}\left(q^2\right) = 
\nonumber \\
&& \hspace{-1.5em} \frac{g^2}{16\pi^2} \int_0^1dx\int_0^{1-x} dy
 \biggl\{
 \frac{1}{\epsilon}-\gamma-\frac{3}{2} + \ln\left(\frac{\mu^2}{\pi M^2_{\rm cov}}\right)
 \biggr.
\nonumber \\
&& \hspace{-1.5em} +
\biggl.
\frac{\left[ m + \left(1-x-y\right)M \right]^2 + xy q^2}{M^2_{\rm cov}}
\biggr\},
\\
&&\hspace{-2.5em} F_2^{\mathsf{scalar}}\left(q^2\right) =  
\nonumber \\ 
&& \hspace{-1.5em} \frac{g^2}{8\pi^2} \int_0^1 dx \int_0^{1-x} dy
 \frac{(x+y)\left[ m + (1-x-y) M\right] M}{M^2_{\rm cov}}. \nonumber
\end{eqnarray}
This is exactly what we get from the standard calculation, i.e.,
${\mathsf N}^\mu=\left(\slash \!\!\!p_2+m \right)\gamma^\mu
\left(\slash \!\!\!p_1+m \right)$ in Eq.~(\ref{current}).

Another example is the calculation of the electron vertex correction
in four-dimensional QED, where the exchanged
particle is a vector photon (i.e $m_X = 0)$. 
In this case, the external fermion lines have the same mass as the internal ones,
i.e. $M=m$. From the Fierz transformation relations
in Table~\ref{t1} for this case, we have: ${\mathsf C}_{\rm S}=1$,
${\mathsf C}_{\rm V}=-\frac{1}{2}$, ${\mathsf C}_{\rm T}=0$, ${\mathsf
C}_{\rm A}=-\frac{1}{2}$, ${\mathsf C}_{\rm P}=-1$.  Substituting these
values in Eqs.~(\ref{4F26}) and~(\ref{4F27}), we get:
\begin{eqnarray}
\label{QEDF1F2}
&&\hspace{-2em} F_1^{\mathsf{QED}}\left(q^2\right) = 
\nonumber \\
&&\hspace{-1em} -
\biggl.
\frac{g^2}{8\pi^2} \int_0^1dx\int_0^{1-x} dy
 \biggl\{
 \frac{1}{\epsilon}-\gamma-\frac{3}{2}
+ \ln\left(\frac{\mu^2}{\pi M^2_{\mathsf{QED}}}\right)
\biggr.
\nonumber \\
&& \hspace{-1em} +
\biggl.
\frac{\left[ (x+y)^2 + 2(x+y-1) \right]m^2 +
 (1-x)(1-y) q^2}{M^2_{\mathsf{QED}}}
\biggr\},
\nonumber\\
&& \hspace{-2em} F_2^{\mathsf{QED}}\left(q^2\right)  = 
\nonumber \\
&& \hspace{-1em}
-\frac{g^2}{4\pi^2} \int_0^1 dx \int_0^{1-x} dy
 \frac{(x+y)(1-x-y)m^2}{M^2_{\mathsf{QED}}}.
\end{eqnarray}
where $M^2_{\mathsf{QED}}=M^2_{\rm cov}(M\to m, m_X \to 0)$.  Again, this is exactly 
the result we get from the standard calculation in Feynman gauge, i.e.,
${\mathsf N}^\mu=\gamma^\nu \left(\slash \!\!\!p_2+m \right)\gamma^\mu
\left(\slash \!\!\!p_1+m \right)\gamma_\nu$ in Eq.~(\ref{current}).

\end{document}